\documentclass[12pt]{article}

\usepackage{hyperref}
\usepackage{amsmath,amsfonts,amssymb}
\hypersetup{dvips,dvipdfm,colorlinks=true,urlcolor=magenta,filecolor=magenta,linktoc=page,citecolor=red,linkcolor=blue,bookmarks=true}
\usepackage{graphicx,epstopdf}
\usepackage{multirow}
\usepackage{xcolor}
\date{}
\begin{document}
	\title{Raychaudhuri equation and Bouncing cosmology}
	\author{Madhukrishna Chakraborty \footnote{chakmadhu1997@gmail.com}~~and~~
		Subenoy Chakraborty\footnote{schakraborty.math@gmail.com (corresponding author)}
		\\Department of Mathematics, Jadavpur University, Kolkata - 700032, India}
	\maketitle
	\begin{abstract}
		The present work deals with an exhaustive study of bouncing cosmology in the background of homogeneous and isotropic Friedmann-Lemaître-Robertson-Walker space-time. The geometry of the bouncing point has been studied extensively and used as a tool to classify the models from the point of view of cosmology. Raychaudhuri equation (RE) has been furnished in these models to classify the bouncing point as regular point or singular point. Behavior of time-like geodesic congruence in the neighbourhood of the bouncing point has been discussed using the Focusing Theorem which follows as a consequence of the RE. An analogy of the RE with the evolution equation for a linear harmonic oscillator has been made and an oscillatory bouncing model has been discussed in this context.
	\end{abstract}
	\section{Introduction}
	In the recent times, there has been an increasing inclination of the cosmologists towards the cosmological models that replace the cosmological singularity (or big bang) with a ``big bounce"---a smooth transition from contraction to expansion in order to resolve fundamental issues in cosmology. The cosmological evolution in the early universe is usually described by the standard Big-Bang cosmology. However the standard cosmology suffers from a couple of issues such as the Horizon problem, Flatness problem, baryon asymmetry and initial singularity. Although inflationary scenario \cite{Guth:1980zm}, successfully resolved some of the issues related to early universe standard model, yet it suffers from the singularity problem and hence fails to reconstruct the complete past history of the universe.\\
	$~~~~~~~~~~~~~~~~~~~~~~~~~~~~~~~~~~~$ At singularity, all physical laws break down. From a geometric point of view singularity can be defined in terms of geodesic incompleteness. Hawking and Penrose used this notion of geodesic incompleteness in the proof of the seminal singularity theorems \cite{Hawking:1973uf}-\cite{Hawking:1970zqf}. It is quite interesting to note that the Raychaudhuri Equation which is an evolution equation for the expansion scalar of a congruence of time-like/ null  geodesics is the main ingredient behind the singularity theorems. The RE together with some conditions on matter proves the inevitable existence of singularity in Einstein gravity. One may refer to \cite{Raychaudhuri:1953yv}-\cite{Chakraborty:2023ork} for a detailed study of the RE. Although after the detection of gravitational waves Einstein's General Theory of Relativity is the most well accepted theory of gravity to describe physical reality yet the main drawback of this theory lies in the inherent existence of singularity. Resolution of this singularity has remained a puzzle over the decades. An alternative model to the standard big bang scenario, describing inflation withour initial singularity had been proposed. This model is known as the Emergent Universe model \cite{Ellis:2002we}. Another non-singular approach without inflation known as the matter bounce scenario have been proposed \cite{Bars:2011mh}-\cite{Brandenberger:2009jq}. For further information regarding non-singular models, one may refer \cite{Brandenberger:2016vhg}-\cite{Novello:2008ra}.\\
	$~~~~~~~~~~~~~$ The initial singularity occurring in the standard Big Bang cosmology and the inflationary cosmology can be suitably avoided in the matter bounce scenario. Bouncing cosmologies have been investigated in extended theories of gravity such as $f(R)$ theory \cite{Carroll:2003wy}-\cite{Ilyas:2022gul}, modified Gauss-Bonnet gravity \cite{Bamba:2014mya}-\cite{Li:2007jm}, $f(R,T)$ gravity \cite{Singh:2018xjv}-\cite{Harko:2011kv}, $f(Q,T)$ gravity \cite{Agrawal:2021rur} and $f(T)$ gravity \cite{Cai:2011tc}. Resolution of the initial singularity by applying Loop Quantum gravity approach gives rise to a hybrid cosmological scenario of the emergent and bouncing universe picture \cite{Li:2023dwy}. In the present work bouncing cosmology has been studied from RE point of view. The work gives light on a plethora of bouncing models and the nature of the bouncing points.\\  The plan of the paper is as follows: Section 2 deals with a brief overview of the Raychaudhuri equation and Focusing theorem in General Relativity. Section 3 gives a clear picture of different bouncing cosmological models. In Section 4, Bouncing models have been studied in the context of the Raychaudhuri equation. An analogy between the Raychaudhuri equation and evolution equation of a classical harmonic oscillator has been shown in Section 5 and an oscillatory bouncing model has been investigated using the harmonic oscillator. In Section 6, nature of bouncing points has been investigated from mathematical point of view. Finally the paper ends with conclusion in Section 7.
	\section{Raychaudhuri equation and Focusing condition in General Relativity}
	The Raychaudhuri equation \cite{Raychaudhuri:1953yv}-\cite{Chakraborty:2023ork} for a congruence of time-like curves having velocity vector field $u^a$ is given by
	\begin{equation}
		\dfrac{d\Theta}{d\tau}=-\dfrac{\Theta^{2}}{3}-2\sigma^{2}+2\xi^{2}+\nabla_{b}A^{b}-R_{ab}u^{a}u^{b}\label{eq1}
	\end{equation} where $\Theta=\nabla_{a}u^{a}$ is the expansion scalar (trace part of the covariant derivative of the vector field $\mathbf{\nabla u}$) ; $2\sigma^{2}=\sigma_{ab}\sigma^{ab}$ and $\sigma_{ab}=\frac{1}{2}(\nabla_{a}u_{b}+\nabla_{b}u_{a})-\frac{1}{3}q_{ab}\Theta+\frac{1}{2}(A_{b}u_{a}+A_{a}u_{b})$ is the shear tensor (trace-less symmetric part of $\mathbf{\nabla u}$), $q_{ab}=g_{ab}+u_{a}u_{b}$ is the induced metric ; $2\xi^{2}=\xi_{ab}\xi^{ab}$ and $\xi_{ab}=\frac{1}{2}(\nabla_{b}u_{a}-\nabla_{a}u_{b})-\frac{1}{2}(A_{b}u_{a}-A_{a}u_{b})$ is the rotation/ vorticity tensor (anti-symmetric part of $\mathbf{\nabla u}$);  $A^{a}=u^{b}\nabla_{b}u^{a}$ is the acceleration vector field and $u_au^{a}=-1$. $R_{ab}$ is the Ricci tensor projected along the time-like curves. The last term on the R.H.S of the Raychaudhuri equation i.e. $-R_{ab}u^{a}u^{b}$ highlights the contribution of space-time geometry and it does not depend on the derivative of the vector field. Therefore this term has more general implications than the other terms in (\ref{eq1}). Geometrically, this term can be interpreted as a mean curvature in the direction of $\textbf{u}$. One may note that (\ref{eq1}) reduces to much simpler form if we assume:\\
	(i)  Congruence of time-like curves to be geodesics. Then along the geodesics we have $A_{a}=0$.\\
	(ii) Congruence of time-like geodesics to be hyper-surface orthogonal which by virtue of Frobenius Theorem implies zero vorticity i.e. $\xi^{2}=0$. Further our aim is to study focusing condition of geodesics as a consequence of RE, hence congruences with vanishing vorticity should be taken into consideration in order to avoid the presence of centrifugal forces.\\
	Under these two assumptions the RE becomes
	\begin{equation}
		\dfrac{d\Theta}{d\tau}=-\dfrac{\Theta^{2}}{3}-2\sigma^{2}-R_{ab}u^{a}u^{b}\label{eq2}
	\end{equation} The RE is a geometric identity in the Riemannian geometry and has nothing to do with gravity. However the role of gravity comes into picture through the last term on the R.H.S of equation (\ref{eq2}) or more specifically through $R_{ab}$.
	In Einstein gravity if matter satisfies SEC i.e. $T_{ab}u^{a}u^{b}+\dfrac{1}{2}T\geq0$ then $R_{ab}u^{a}u^{b}\geq0$. Hence from (\ref{eq2}) $\dfrac{d\Theta}{d\tau}+\dfrac{\Theta^{2}}{3}\leq0$ which upon integration w.r.t $\tau$ yields the inequality
	\begin{equation}
		\dfrac{1}{\Theta}\geq \dfrac{1}{\Theta_{0}}+\dfrac{\tau}{3}
	\end{equation} This is nothing but the mathematical version of the Focusing theorem (FT) which states that : An initially converging congruence of time-like geodesic (orthogonal to the space-like hyper-surface) begin to develop a caustic within finite value of the proper time. This leads to Focusing of geodesic and the corresponding condition on matter ($R_{ab}u^{a}u^{b}\geq0$) is coined as Convergence Condition (CC). This CC essentially hints at the attractive nature of gravity which leads to geodesic focusing and formation of singularity. Although singularity here is a congruence singularity and may not be a space-time singularity but these conditions together with other global arguments may sometimes lead to cosmological and black-hole singularities. Thus a violation of the CC may lead to possible avoidance of singularity. This tool is exploited in various modified theories of gravity where the field equations differ from the Einstein's field equations and lead to violation of CC ($R_{ab}u^{a}u^{b}<0$) under certain circumstances (for ref see \cite{Choudhury:2021zij}, \cite{Chakraborty:2023ork}). Also this FT gives the clue about the origin of the universe from a big-bang singularity. Most importantly,  FT and RE ultimately proved the inevitable existence of singularity in Einstein gravity as its biggest drawback and are also regarded as the key ingredients behind the seminal singularity theorems by Hawking and Penrose \cite{Hawking:1973uf}-\cite{Hawking:1970zqf}.
	\section{Bouncing Scenario}
	In the present work we examine the consequences of the RE for two types of bouncing models, namely 	$\mathbf{B1}$ and 	$\mathbf{B2}$. The present section deals with the details of the two models as follows:
	\section*{$\underline{\mathbf{B1}}$: Bouncing Point is a local minima for $a(t)$}
\begin{figure}
		\includegraphics[height=4cm,width=6cm]{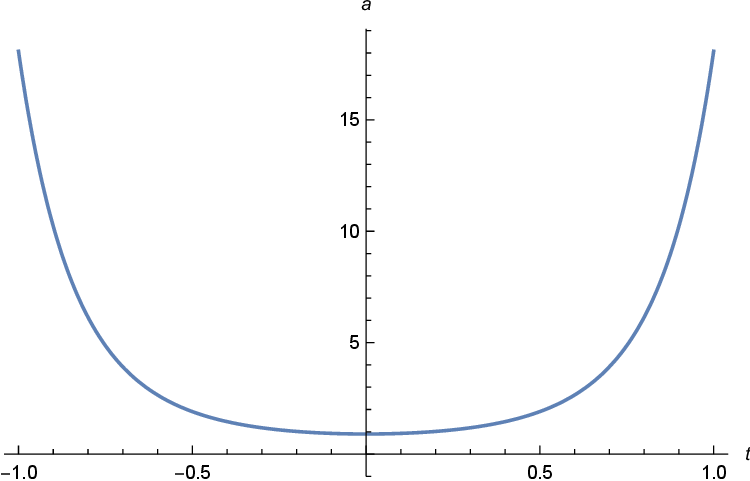}\hfill
		\includegraphics[height=4cm,width=6cm]{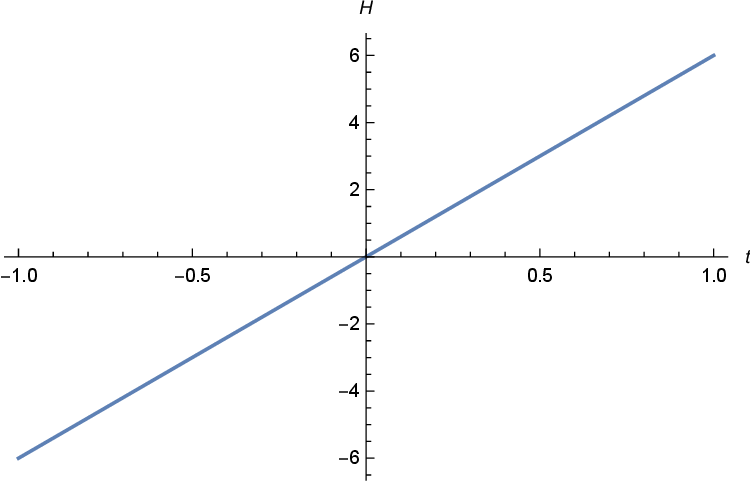}
		\caption{Scale factor $a(t)$ vs $t$ (left) and Hubble parameter $H$ vs $t$ (right) representing example 1 of $\mathbf{B1}$}\label{f1}
	\end{figure}

\begin{figure}
		\includegraphics[height=4cm,width=6cm]{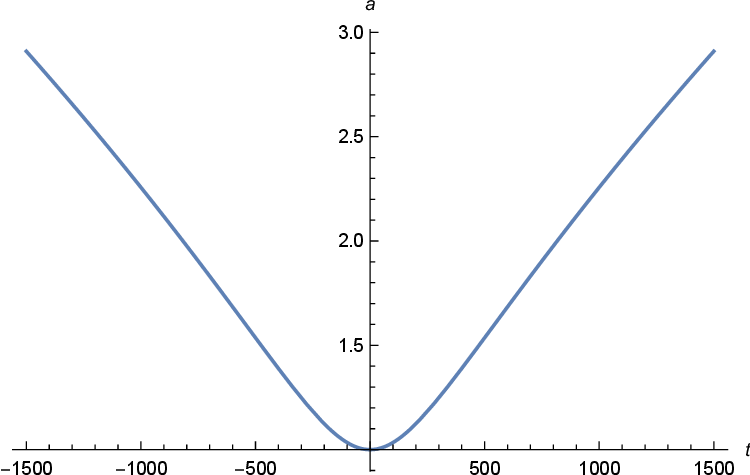}\hfill
		\includegraphics[height=4cm,width=6cm]{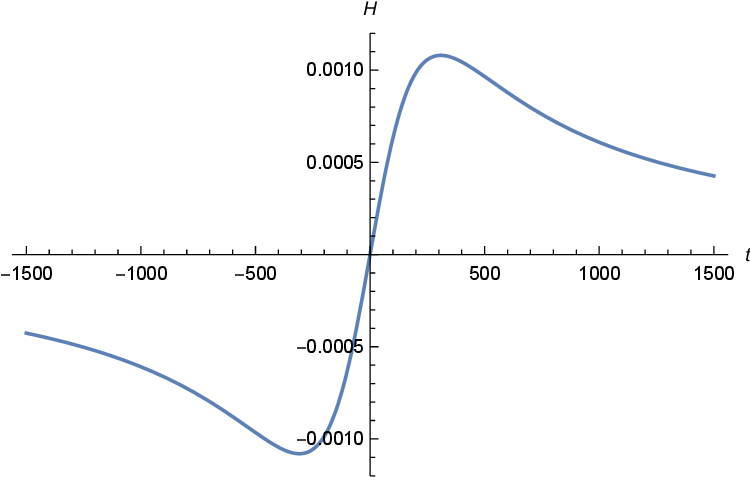}
		\caption{Scale factor $a(t)$ vs $t$ (left) and Hubble parameter $H$ vs $t$ (right) representing example 2 of $\mathbf{B1}$}\label{f2}
	\end{figure}

	\textbf{Behavior of cosmic scale factor $a(t)$ and Hubble parameter $H$:}
	\begin{itemize}
		\item $a(t)$ decreases before bounce, attains a minimum at the bouncing point and then increases after the bounce. This is illustrated in figures \ref{f1}-\ref{f4} (The variation of scale factor is shown)
		\item $H<0$ before bounce, $H=0$ at the bouncing point and $H>0$ after the bounce.
		\item Continuity of $H$ can be shown graphically in figures \ref{f1}-\ref{f4}  which clearly depict that $H$ is an increasing function i.e.  $\dot{H}>0$ in the deleted neighbourhood of bounce.
	\end{itemize}
	Some examples of $\mathbf{B1}$ are:
	\begin{enumerate}
		\item Symmetric Bounce \cite{Cai:2012va}: $a(t)=B\exp\left(\beta~\dfrac{t^{2}}{t_0^{2}}\right)$, $H=\dfrac{2\beta t}{t_0^{2}}$,  $t_0$ is some arbitrary time $B>0$, $\beta>0$.
		\item Matter Bounce \cite{Singh:2006im}, \cite{Wilson-Ewing:2012lmx}:
		$a(t)=A\left(1.5\rho_ct^{2}+1\right)^{\frac{1}{3}}$, $H=\dfrac{2\rho_ct}{3\rho_ct^{2}+2}$ where $A>0$ is a constant, $0<\rho_c<<1$ is the critical density whose value stems from LQC.
		\item Type I-IV (past/future) singularities and little rip cosmologies \cite{Caruana:2020szx}:  $a(t)=A\exp\left[\dfrac{f_0}{\alpha+1}(t-t_s)^{\alpha+1}\right]$, $H=f_0(t-t_s)^\alpha$, $\alpha\neq-1,0,1$ and $\alpha$ must be odd. $t_s$ is the time at which bounce occurs.
		\item A bounce characterized by $a(t)=b\exp(k\exp(t-\alpha)-t)$, $H=k(\exp(t-\alpha)-1)$~~($k, \alpha,b>0)$
	\end{enumerate}
	\begin{figure}
		\includegraphics[height=4cm,width=6cm]{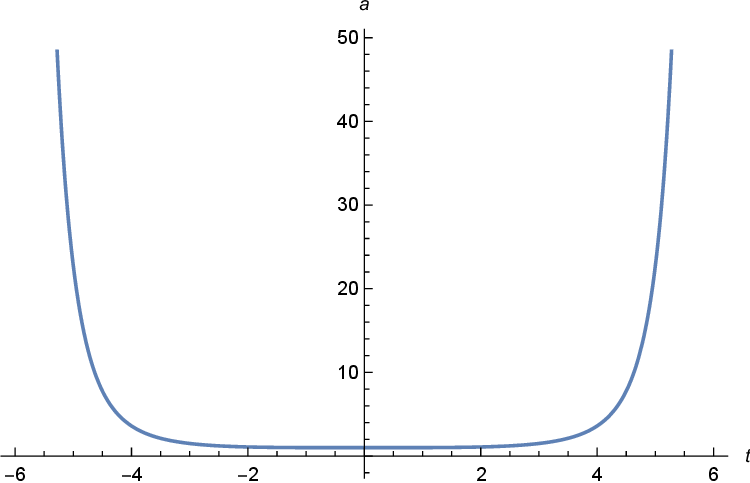}\hfill
		\includegraphics[height=4cm,width=6cm]{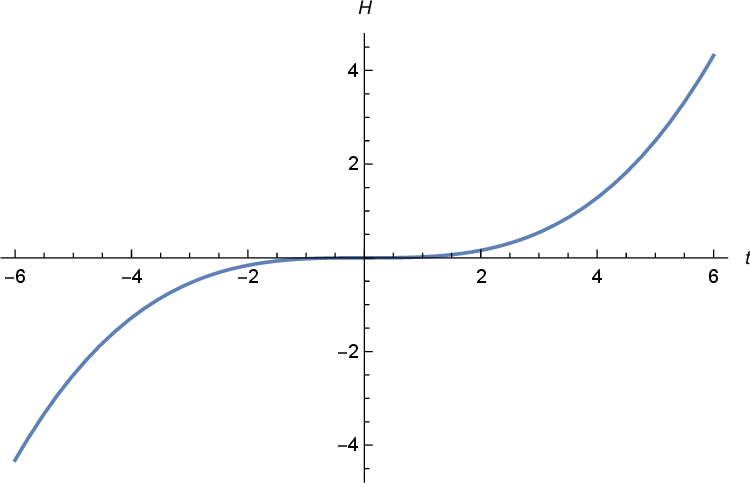}
		\caption{Scale factor $a(t)$ vs $t$ (left) and Hubble parameter $H$ vs $t$ (right) representing example 3 of $\mathbf{B1}$}
	\end{figure}\label{f3}
	\begin{figure}
		\includegraphics[height=4cm,width=6cm]{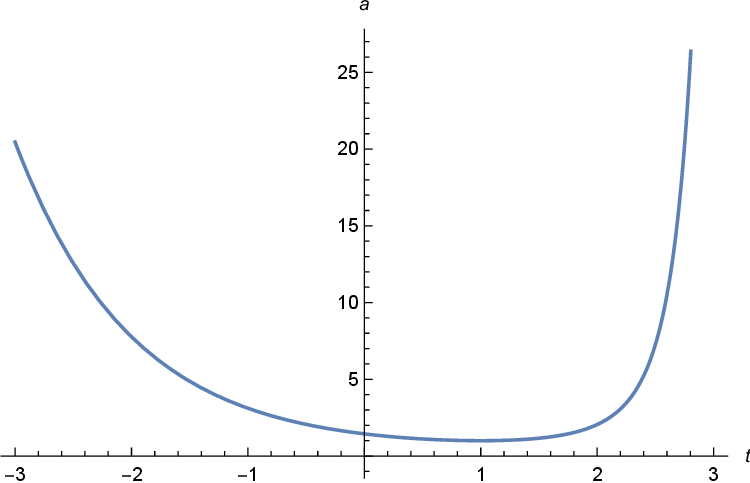}\hfill
		\includegraphics[height=4cm,width=6cm]{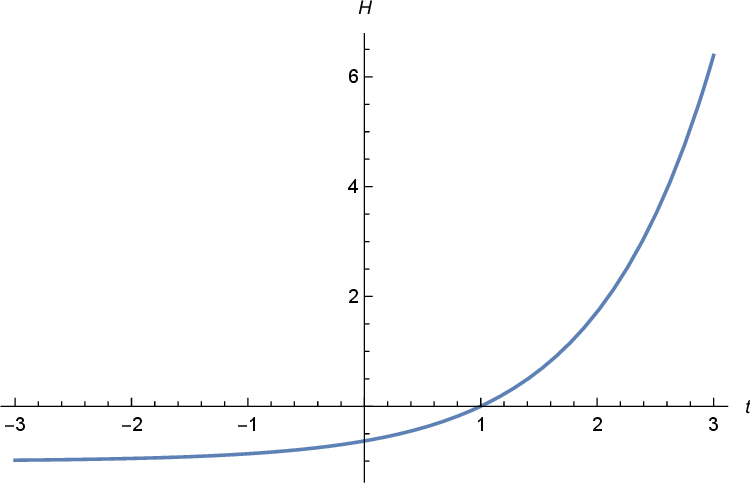}
		\caption{Scale factor $a(t)$ vs $t$ (left) and Hubble parameter $H$ vs $t$ (right) representing example 4 of $\mathbf{B1}$}\label{f4}
	\end{figure}
	\section*{$\underline{\mathbf{B2}}$ : Bouncing Point is a local maxima for $a(t)$}
\begin{figure}
		\includegraphics[height=4cm,width=6cm]{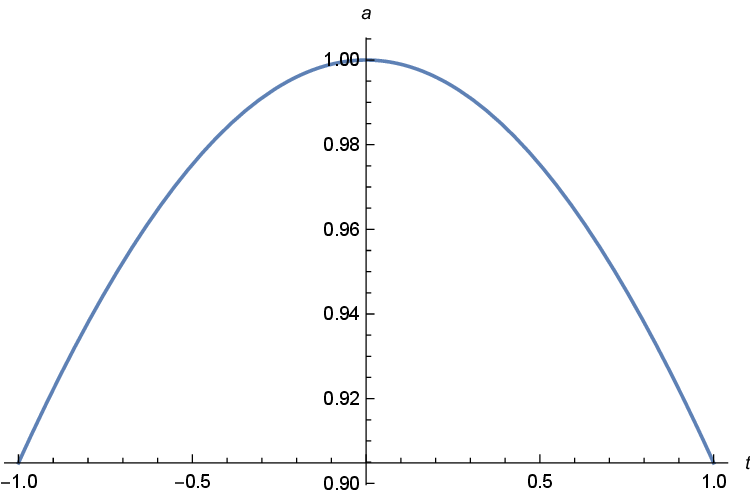}\hfill
		\includegraphics[height=4cm,width=6cm]{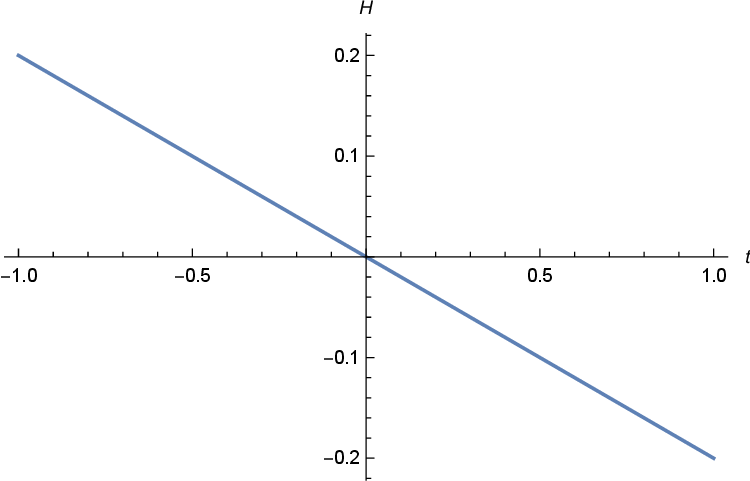}
		\caption{Scale factor $a(t)$ vs $t$ (left) and Hubble parameter $H$ vs $t$ (right) representing a case of $\mathbf{B2}$}\label{f5}
	\end{figure}

\begin{figure}
		\includegraphics[height=4cm,width=6cm]{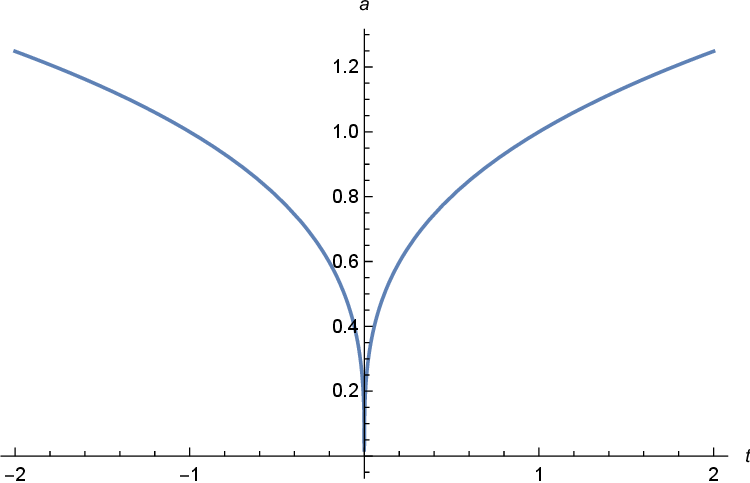}\hfill
		\includegraphics[height=4cm,width=6cm]{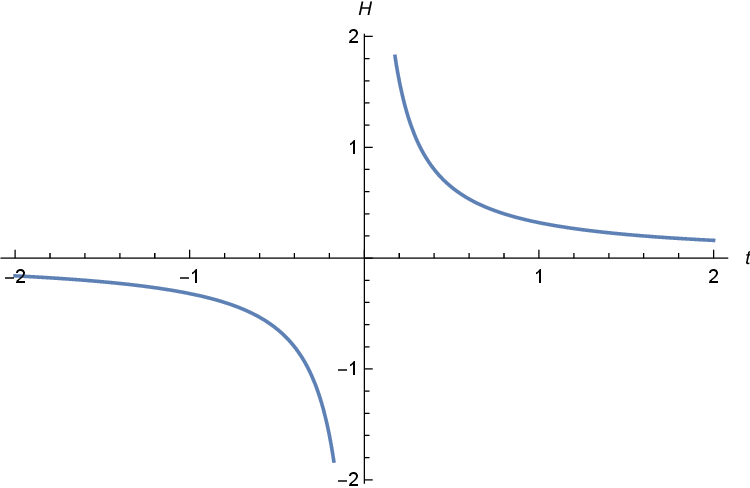}
		\caption{Scale factor $a(t)$ vs $t$ (left) and Hubble parameter $H$ vs $t$ (right) for superbounce 1.1}\label{f6}
	\end{figure}

\begin{figure}
		\includegraphics[height=4cm,width=6cm]{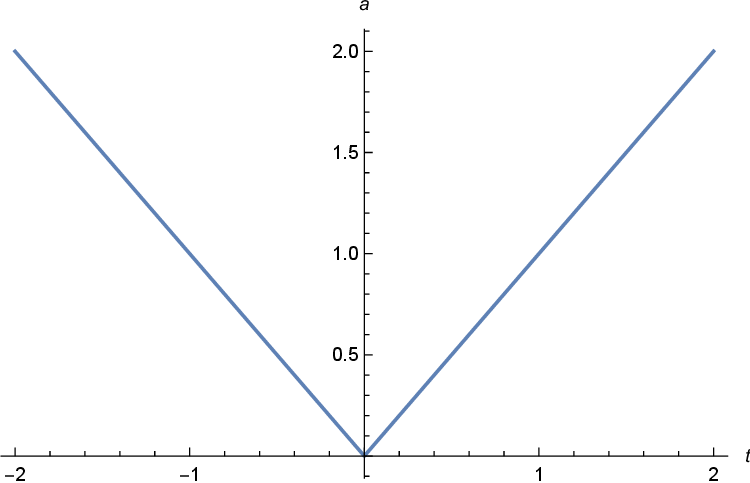}\hfill
		\includegraphics[height=4cm,width=6cm]{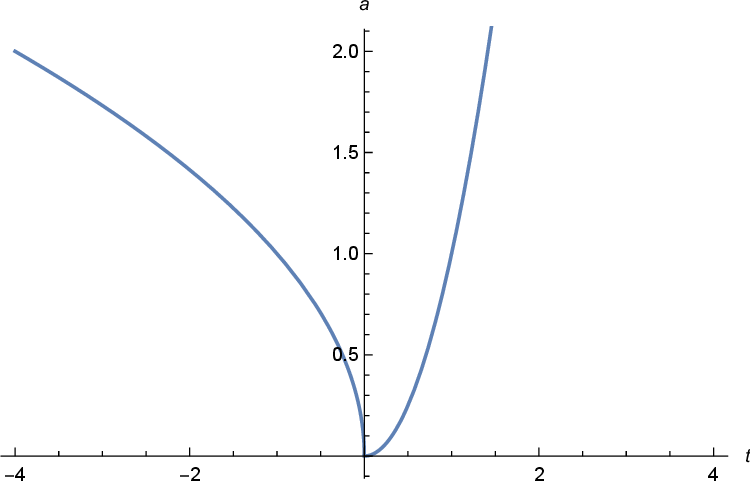}
		\caption{Scale factor $a(t)$ vs $t$ for the bounce 1.2 (left) and 1.3 (right)}\label{f7}
	\end{figure}

	\textbf{Behavior of cosmic scale factor $a(t)$ and Hubble parameter $H$:}
	\begin{itemize}
		\item $a(t)$ increases before bounce, attains a maximum at the bouncing point and then decreases after the bounce. This is illustrated in Fig \ref{f5} (an example).
		\item $H>0$ before bounce, $H=0$ at the bouncing point and $H<0$ after the bounce.
		\item Continuity of $H$ can be shown graphically in Fig \ref{f5} which clearly depicts that $H$ is a decreasing function i.e.  $\dot{H}<0$ in the deleted neighbourhood of bounce.
	\end{itemize}
	Example: A bounce characterized by $a(t)=A\exp\left( b\left(\dfrac{t}{t_0}\right)^2\right)$, $H=\dfrac{2bt}{t_0^2}$ where $A>0$, $b<0$ $t_0>0$ is an arbitrary time. This bounce is illustrated in Fig \ref{f5}.\\
	\\
	\textbf{Some special cases:}
	\begin{itemize}
		\item 1.1 Superbounce \cite{Koehn:2013upa}, \cite{Odintsov:2015uca}: $a(t)=\left(\dfrac{t_s-t}{t_0}\right)^{\frac{2}{c^{2}}}$, $H=-\dfrac{2}{c^{2}}\left(\dfrac{1}{t_s-t}\right)$ where $c>\sqrt{6}$ is a constant, $t_s$ is the time at which bounce occurs. This bouncing model is shown in Fig \ref{f6}. 
		\item 1.2 Bounce characterized by scale factor $a(t)=|t|$. This bounce is shown in Fig \ref{f7} (left).
		\item 1.3 Bounce characterized by 
		$$a(t)=	  \left\{
		\begin{array}{ll}
			\sqrt{-t} & t\leq0 \\
			t^{2} & t\geq0 \\
		\end{array} 
		\right. $$
	\end{itemize} ~~~~~~~~~~This bounce is shown in Fig \ref{f7} (right).
	\section{Raychaudhuri equation in FLRW model and bouncing scenario}
	For Friedmann–Lemaître–Robertson-Walker (FLRW) model,  $\Theta=3H=3\frac{\dot{a}}{a}$ and $\sigma^{2}=0$. Thus the RE (\ref{eq2}) takes the form
	\begin{equation}
		3\dot{H}=-3H^{2}-R_{ab}u^{a}u^{b}.\label{eq3}
	\end{equation} In the context of cosmology $\tau$ can be treated as the cosmic time $t$ and `.' represents differentiation w.r.t $t$.\\
	\textbf{\underline{RE and B1}:} In the previous section we have seen that $\dot{H}>0$ in the deleted neighbourhood of the bouncing point in $\textbf{B1}$. Therefore from the RE (\ref{eq3}) one has\\
	(i) $R_{ab}u^{a}u^{b}<0$ and, \\
	(ii) $|R_{ab}u^{a}u^{b}|>3H^{2}$ in the deleted neighbourhood of $\textbf{B1}$. \\If the matter content of the universe is a perfect fluid having energy density $\rho$ and pressure $p$ then  $R_{ab}u^{a}u^{b}=\dfrac{1}{2}(\rho+3p)$. The RE in FLRW space-time is given by
	\begin{equation}
		\dfrac{\ddot{a}}{a}=-\dfrac{4}{3}(\rho+3p)\label{eq5*}
	\end{equation} Thus using (i) in equation (\ref{eq5*}) one may conclude that in this bouncing model there is always acceleration. Violation of SEC indicates that matter is exotic in nature and singularity can be avoided near the bounce (in the sense that CC needed for focusing of geodesic congruence is violated here). Further in this type of bounce, the behavior of energy density $\rho$ can be studied using the RE. From the first Friedmann equation we have $3H^{2}=\rho$ and from the RE (\ref{eq3}) we have $2\dot{H}=-(\rho+p)$ which is nothing but the second Friedmann-equation. Hence for $\textbf{B1}$ in the deleted neighbourhood of bounce $(\rho+p)<0$ i.e. Null Energy Condition (NEC) is violated. To study the continuity of $\rho$ we shall use the matter conservation equation 
	\begin{equation}
		\dot{\rho}+3H(p+\rho)=0\label{eq4}
	\end{equation} Thus before bounce as $H<0$ so $\dot{\rho}<0$. At bouncing point $H=0$ and hence from the first Friedmann equation $\rho=0$. After bounce $H>0$ and hence $\dot{\rho}>0$. Thus continuity of $\rho$ can be studied as: $\rho$ is decreasing before bounce, attains zero value at bouncing point and increases after bounce. If we assume further that the matter content of the universe is a perfect fluid with barotropic equation of state $p=\omega\rho$, $\omega$ being the equation of state (EoS) parameter then in the neighbourhood of $\textbf{B1}$, $\omega<-1$ i.e. phantom energy favors this type of bounce. From the second Friedmann equation $\dot{H}=0$ at the bouncing point. But $q$, the deceleration parameter is not defined at the bouncing point. Although this type of bounce can avoid the initial big-bang singularity as evident from the RE, yet the peculiarity observed at the bouncing point hints that the bouncing point may be regarded as a higher order singularity in the sense that $a$, $H$, $\dot{H}$ are defined at the bouncing point but no other higher order parameters like $q$ and $j$ are defined at the bouncing point. However there is always acceleration in this type of model except at the bouncing point where $\ddot{a}=0$.\\ \\
	\textbf{\underline{RE and B2}:} For $\textbf{B2}$, $\dot{H}<0$ in the  neighbourhood of the bouncing point and $H=0$ at the bouncing point. Therefore from the RE (\ref{eq3}) there are two possibilities: (i) $R_{ab}u^{a}u^{b}>0$ or (ii) $R_{ab}u^{a}u^{b}<0$ but $|R_{ab}u^{a}u^{b}|<3H^{2}$ in the neighbourhood of the bouncing point in $\textbf{B2}$. 
	\begin{enumerate}
		\item In the first case convergence condition (CC) or precisely the Strong Energy Condition (SEC) holds. Therefore bounce occurs even with normal/usual matter. There is always deceleration except at the bouncing epoch as clear from the RE (\ref{eq5*}). Since $a_{b}\neq0$, hence at the bouncing point there is non-zero volume (hence no singularity). However focusing/convergence condition holds in the neighbourhood of bounce. This shows focusing alone does not always leads to singularity formation. However the converse is true i.e, if there is a singularity of the space-time then a congruence of time-like/ null geodesic will focus there.
		\item  For the second case bounce occurs with exotic matter having EoS parameter $-1<\omega<-\frac{1}{3}$ and violation of CC results in the avoidance of big-bang singularity. There is always acceleration except at the bouncing point where $\ddot{a}=0$.
	\end{enumerate} The energy density $\rho$ shows a similar behavior just like in $\textbf{B1}$. $\dot{H}=0$ at the bouncing point. In $\textbf{B2}$ also $a$, $H$ and $\dot{H}$ are defined at the bouncing point but next higher order derivatives are undefined at the bouncing point. This absurdness of the bouncing point again hints the existence of 2nd order singularity. Therefore the RE essentially depicts the existence of higher order singularities at the bouncing point.\\
	~~~~~~~~~~~~~~~The full analysis done till now has been carried out for $\kappa=0$. Now let us investigate which value of $\kappa$ favors the bouncing scenario. For $\kappa\neq0$, the Einstein's field equations are
	\begin{eqnarray}
		3\left(H^{2}+\frac{\kappa}{a^{2}}\right)=\rho\label{eq5}\\
		2\left(\dot{H}-\frac{\kappa}{a^{2}}\right)=-(\rho+p)\label{eq6}
	\end{eqnarray} Since $H^{2}>0$, from (\ref{eq5}) $\rho>\frac{3\kappa}{a^{2}}$.  For $\kappa=+1$, $\rho$ is positive definite but for $\kappa=-1$, $\rho=3\left(H^{2}-\frac{1}{a^{2}}\right)$, therefore one must have a lower bound on H i.e. $H^{2}>\dfrac{1}{a^{2}}$ in order to have a normal matter in the neighbourhood of bouncing point. More specifically, $H>\dfrac{1}{a}$ after the bounce and $H<-\dfrac{1}{a}$ before the bounce i.e. $\dot{a}>1$ after bounce, $\dot{a}<-1$ before bounce and $\dot{=0}$ at the bouncing point. At the bouncing point $H=0$ which implies $\rho_{b}=\frac{3\kappa}{a_{b}^{2}}\neq0$ ($\rho_{b}$ and $a_{b}$ are the energy density and scale factor at the bouncing epoch $t_{b}$). Further it is to be noted that for $\kappa=-1$ energy density at bouncing point ($\rho_{b}$) is negative which says that matter must be ghost type at the bouncing point. From (\ref{eq6}), $\dot{H}=\dfrac{\kappa}{a^{2}}-\dfrac{1}{2}(\rho+p)$.  For $\textbf{B1}$, $\dot{H}>0\implies\rho+p<\frac{2\kappa}{a^{2}}$. So for $\kappa=-1$, the matter is of phantom type, while for $\kappa=+1$, the matter may not be exotic in nature. Therefore the above analysis shows that for bouncing scenario $\kappa=+1$ is more suitable than $\kappa=-1$.\\
	~~~~~~~~~~~~~~~~~~~~For $\kappa=0$ we have seen that perfect fluid with barotropic equation of state can not give the true picture of a bouncing model. This is because we do not have any idea about the behavior of $q$, $j$ and other parameters involving higher derivatives of $H$. Therefore let us choose two different types of equation of state as follows:\\
	\textbf{(i) van der Waals  equation of state} : The EoS is given by
	\begin{equation}
		p=\dfrac{A\rho}{1-B\rho}-C\rho^{2}
	\end{equation} where $A$, $B$ and $C$ are constants. So $\omega=\dfrac{p}{\rho}=\dfrac{A}{1-B\rho}-C\rho\rightarrow A$ as $\rho\rightarrow0$  i.e. at the bouncing point $\omega=A$, a constant hence defined. Also $2\dot{H}=-(\rho+p)=-\left(\rho+\dfrac{A\rho}{1-B\rho}-C\rho^{2}\right)$ and $3H^{2}=\rho$. $\therefore q=-(1+\frac{\dot{H}}{H^{2}})=\dfrac{(1+3A)}{2}$. Therefore $q$ is defined at the bouncing point for these bouncing models with perfect fluid having van der waals EoS. Further the sign of $q$ depends on $A$. Thus acceleration (deceleration) near bounce is determined by $A$.\\
	\textbf{(ii) Polytropic equation of state}: The EoS is given by
	\begin{equation}
		p=k\rho^{(1+\frac{1}{n})}
	\end{equation} where $k$ is a proportionality constant and $n$ is the polytropic index ($n$ is any real number). Here $\omega=\dfrac{p}{\rho}=k\rho^{\frac{1}{n}}\rightarrow0$ as $\rho\rightarrow0$ at the bouncing point for $n>0$ and $q=\dfrac{1}{2}$ at the bouncing point indicating deceleration near bounce. However for $n\leq0$ both $q$ and $\omega$ are undefined. Hence it shows that $q$ and $\omega$ are defined at bouncing point of these bouncing models with perfect fluid having polytropic EoS with a positive polytropic index.
	\section{Raychaudhuri equation: A Linear Harmonic Oscillator and Oscillatory bounce}
\begin{figure}
		\includegraphics[height=5cm,width=6cm]{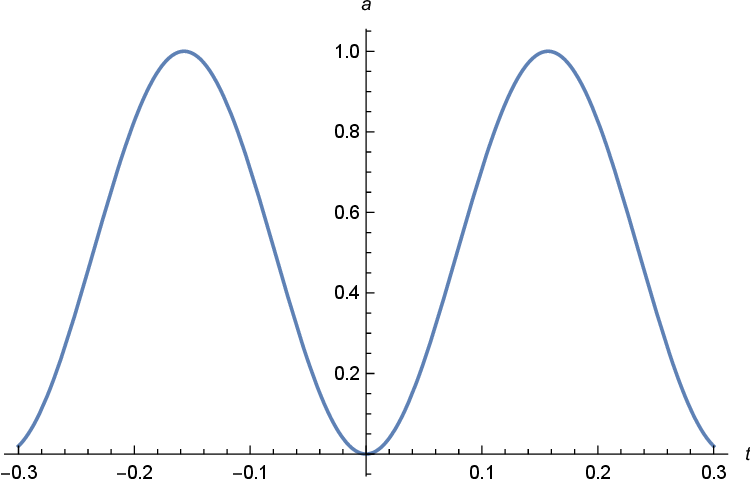}\hfill
		\includegraphics[height=5cm,width=6cm]{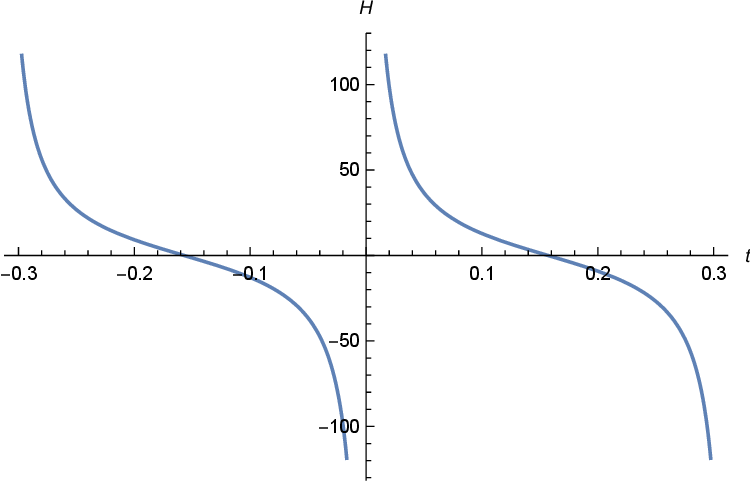}
		\caption{Scale factor $a(t)$ vs $t$ (left) and Hubble parameter $H$ vs $t$ (right) representing an oscillatory bounce.}\label{f8}
	\end{figure}

\begin{figure}
		\centering	\includegraphics[height=8cm,width=10cm]{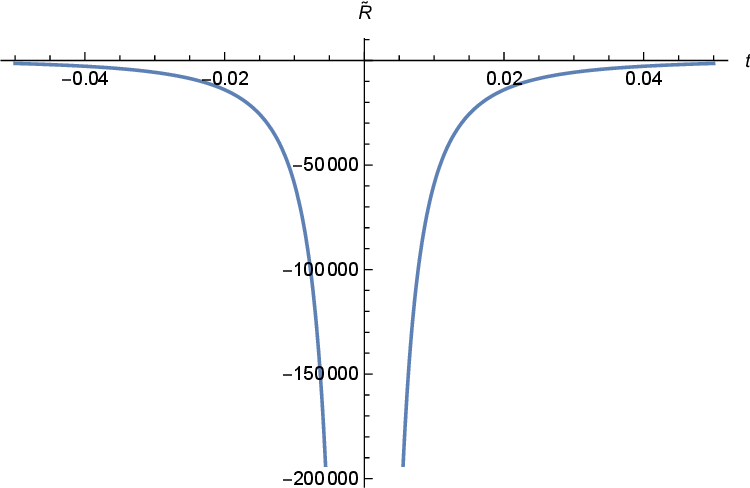}
		\caption{Variation of $\tilde{R}$ with cosmic time $t$ for the oscillatory bouncing cosmological model.}\label{f9}
	\end{figure}

	In this section, we aim to explore whether/how the Raychaudhuri equation can explain the existence and avoidance of initial big-bang singularity in case of oscillatory bouncing cosmological model. To do so we recall the general form of the RE in (\ref{eq1}) and look at the RE from the point of view of an evolution equation for a real harmonic oscillator. Geometrically $R_{ab}u^{a}u^{b}=\tilde{R}$ can be interpreted as mean curvature in the direction of $u$ \cite{Albareti:2012se}. From mathematical point of view the RE can be termed as Riccati equation and it becomes a linear second order equation as 
	\begin{equation}
		\dfrac{d^{2}Y}{d\tau^{2}}+\omega_{0}^{2}~Y=0,
	\end{equation} where 
	\begin{equation}
		\Theta=(n-1)\dfrac{d}{d\tau}\ln Y
	\end{equation} and 
	\begin{equation} \omega_{0}^{2}=\dfrac{1}{n-1}(\tilde{R}+2\sigma^{2}-2\omega^{2}-\nabla_{b}A^{b}).
	\end{equation} Thus the RE can be identified as a linear harmonic oscillator equation with time varying frequency $\omega_{0}$. As 	$\Theta$ may be defined as the derivative of the geometric entropy ($S$) or an average (or effective)
	geodesic deviation so one may identify $S = \ln Y$. The expansion $\Theta$ is nothing but the rate of change of volume of the transverse subspace of the congruence/bundle of geodesics. Therefore, the expansion approaching negative infinity (i.e. $\Theta\rightarrow-\infty$) implies a convergence of
	the bundle, whereas a value of positive infinity (i.e. $\Theta\rightarrow+\infty$) would imply a complete divergence. Thus the
	Convergence Condition (CC) can be stated as follows:\\
	$~~~~$	(i) Initially $Y$ is positive but decreases with proper time i.e $\dfrac{dY}{d\tau}<0$.\\
	$~~~~~$(ii)  Subsequently $Y=0$ at a finite proper time to have negative infinite expansion.\\
	From the above interrelation: $\Theta=\dfrac{(n-1)}{Y}\dfrac{dY}{d\tau}$, it is clear that there should be an initially negative expansion (i.e. $\Theta(\tau=0)<0$) and subsequently $\Theta\rightarrow-\infty$ as $Y\rightarrow0$ at a finite proper time. Therefore the CC essentially coincides with the condition for the existence of zeroes of $Y$ in finite proper time. However the Sturm comparison theorem (in differential equation) shows that the existence of zeros in $Y$ at finite value of the proper time $\tau$ requires
	\begin{equation}
		(\tilde{R}+2\sigma^{2}-2\omega^{2}-\nabla_{b}A^{b})\geq0\label{eq34*}.
	\end{equation} Therefore (\ref{eq34*}) is the CC for a congruence of time-like curves (may be geodesic or non geodesic). Further, the above inequality shows that, the Raychaudhuri scalar $\tilde{R}$ and the shear/anisotropy scalar $2\sigma^{2}$ are in favor of convergence of the congruence of time-like curves while rotation and acceleration terms oppose the convergence. The CC reduces to $\tilde{R}\geq0$ if we consider the congruence of time-like
	curves to be geodesic and orthogonal to the space-like hyper-surface. This leads to \textbf{Geodesic
		Focusing} and hence the \textbf{Focusing Theorem}. In other words, rotation and acceleration terms act against the focusing but shear and
	Raychaudhuri scalar are in favor of it. Thus from physical point of view, if the RE corresponds to a realistic linear harmonic oscillator then it is inevitable to have a singularity. We name the scalar: $R_{c}=\tilde{R}+2\sigma^{2}$, as the Convergence scalar. In FLRW background $R_{c}=\tilde{R}$. Thus avoidance of singularity may be guaranteed by avoiding the focusing of geodesics or by disregarding the CC.\\ \\
	~~~~~~~~~~~~~~ In the present context, $Y=a(t), ~n=4,~\sigma^{2}=\omega^{2}=\nabla_{b}A^{b}=0,~\tau=t$. The RE in real linear harmonic oscillator form can be written as:
	\begin{equation}
		\ddot{a}=-\dfrac{1}{3}~\tilde{R}~a.\label{eq13}
	\end{equation}
	We now consider an oscillatory bouncing cosmological model \cite{Novello:2008ra} characterized by $a(t)=A~\sin^2\left(\dfrac{C~t}{t_{*}}\right)$, $H=\dfrac{2C}{t_{*}}\cot\left(\dfrac{Ct}{t_{*}}\right)$ where $A,~C>0$, $t_{*}$ is some reference time. For the sake of convenience we choose $t_{*}>0$. This model represents the behavior of a cyclic universe, which treats the universe as a continuous sequence of contraction and expansion. The model is shown graphically in Fig \ref{f8}. For this bounce, the RE (\ref{eq13}) gives the expression of the Raychaudhuri scalar $\tilde{R}$ as:
	\begin{equation}
		\tilde{R}=-6\left(\dfrac{C}{t_{*}}\right)^{2}\dfrac{\cos\left(\frac{2Ct}{t_{*}}\right)}{\sin^{2}\left(\dfrac{Ct}{t_{*}}\right)}.\label{eq14}
	\end{equation}The bounce occurs at $t=0$ and $\tilde{R}$ is singular at the bouncing epoch. However the plot of $\tilde{R}$ vs $t$ shows that in very small neighbourhood of the bouncing point singularity is avoided since $\tilde{R}\leq0$ avoids the focusing of geodesics over there (see Fig \ref{f9}). This type of bounce avoids the geodesic focusing in a small neighbourhood of the bouncing point as evident from the RE. However at the bouncing point $a$ vanishes and $H$ becomes singular. This singularity is experienced throughout each cycle when the scale factor goes to zero. The first bounce occurs ar $t=\dfrac{n\pi t_{*}}{C}$ for $n$, an integer and this corresponds to a Big Crunch singularity. This can be resolved by constructing a non zero scale factor through other mechanisms. The second bounce occurs when the universe reaches its maximal size at $t=\dfrac{(2n+1)\pi t_{*}}{2C}$ for an integer $n$ leading to a cosmological turnaround. This represents the instance when the universe stops expanding and starts to contract towards the Big Crunch singularity.
	\section{Bouncing Point : A point of Inflection/ Cusp/ Corner}
	In this section we examine whether the bouncing point in these models namely $\textbf{B1}$ and $\textbf{B2}$ is a point of inflection or a point of different nature and discuss its consequences wherever applicable/ possible.\\\\
	$\textbf{B1}$: Let $t=t_b$ be the bouncing epoch. In this model from the graph of $H$ vs $t$ (see Fig \ref{f2} and \ref{f3}), it is clear that the curve $H(t)$ changes its concavity at $t=t_{b}$. Hence from the definition of point of inflection we can say that the bouncing epoch is a point of inflection of $H(t)$ in some cases of $\textbf{B1}$, namely matter bounce and Type I-IV (past/future)  singularities and little rip cosmologies. Using the necessary and sufficient condition for a function to have a point of inflection we have the following:
	\begin{enumerate}
		\item $\ddot{H}(t)=0$ at $t=t_{b}$.
		\item $\ddot{H}(t)<0$ for $t<t_{b}$ (if $H$ is concave downwards in $t<t_{b}$) or in other words $\dot{H}$ is decreasing in $t<t_{b}$.
		\item $\ddot{H}(t)>0$ for $t>t_{b}$ (if $H$ is concave upwards in $t>t_{b}$) or in other words $\dot{H}$ is increasing in $t>t_{b}$.
	\end{enumerate}
	The above findings help us to draw the graph of $\dot{H}$ vs $t$ in $\textbf{B1}$ model. For example the above analysis holds good in case of Type I-IV (past/future) singularities and little rip cosmologies (see Fig \ref{f3}). The continuity of $\dot{H}$ can be written as: $\dot{H}$ is decreasing before bounce, attains a zero value at the bouncing point (follows from the RE using $\rho=0$ at bouncing point) and then increases after the bounce. The graph for $\dot{H}$ vs $t$ in case of $\textbf{B1}$ has been shown in Fig . It further shows that $\dot{H}>0$ in the neighbourhood of bounce which is consistent with our analysis of $\textbf{B1}$ in Section II. The bouncing point is a rising point of inflection for the curve $H=H(t)$ in some cases of $\mathbf{B1}$. Similar analysis can be performed in case of $\mathbf{B2}$. \\ \\
	~~~~~~~~~~~~~~~~~~~~~~To sum up, the advantage of getting the bouncing point as a point of inflection for the models lies in the information regarding the continuity of $\dot{H}$ and value of $\ddot{H}$ at the bouncing point. Further from the graph of $\dot{H}$ vs $t$ in $\textbf{B1}$ and $\textbf{B2}$ one can draw the graph of $\rho+p$ vs $t$. This further helps to distinguish the models from the point of view of cosmology.\\
	In Bounce 1.3 we have two distinct directions of the tangent. So $H$ is singular at the bouncing point. Since the curve $a(t)$ changes its concavity across the bouncing point in this case bouncing epoch is a point of inflection w.r.t the curve $a(t)$. But no further analysis can be done due to the singular nature of the Hubble parameter $H$ and hence higher order cosmographic parameters like $j$, $s$ etc.\\ \\
	\textbf{Some cases where the bouncing point is not a point of inflection:}
	\begin{enumerate}
		\item In example 1 i.e, in symmetric bounce the Hubble parameter turns out to be linear. Hence the bouncing point is not a point of inflection in this case. Further in example 4 the variation of Hubble parameter is concave upward throughout. Therefore in this case also the bouncing point fails to be a point of inflection.
		\item In case of Super bounce 1.1, the bouncing point is a cusp. It is an infinitely sharp corner. The vertical line at the bouncing point is the tangent at the bouncing point. On one side derivative is $+\infty$ and on the other side the derivative is $-\infty$. So $H$ does not exist at the bouncing epoch as clear from Fig 6.
		\item For the bounce 1.2, the bouncing point is a corner point. Both left and right hand derivative exist but they are not equal. There are two distinct tangents to the two branches at the bouncing point.
	\end{enumerate}
	\textbf{Both types 1 and 2 have first order singularities at the bouncing point.}\\ \\
	 Finally, though the present work is a theoretical (more specifically mathematical) study of bouncing scenario, still it is worthy to mention some areas regarding whether a bounce can actually occur in practice or what are the challenges faced by these models to be practically relevant.
		In the present work we have considered three general type of bouncing cosmological models namely, $\textbf{B1}$, $\textbf{B2}$ and Oscillatory bouncing cosmological model. Subsequently, we have implemented the Raychaudhuri equation (RE) in these models to deduce the criteria for occurrence of such bounce and hint at the geometry of the bouncing point as regular or singular point. Based on the theoretical formulation of the models we have:
	\begin{enumerate} \item  The theoretical implications of $\textbf{B1}$ demands to break a series of singularity theorems by Hawking and Penrose which uses RE as a key ingredient. Such an issue is accompanied by violation of NEC since we restricted our study within the framework of GR. Thus construction of $\textbf{B1}$ in reality without theoretical pathologies is not easy as the scenario is associated with NEC violation which is again accompanied by quantum instabilities. The cosmological model involving a contraction phase (e.g $\textbf{B1}$) suffers from BKL (Belinsky-Khalatinov-Lifschitz) instability issue. Also the examples of $\textbf{B1}$ are non-singular bouncing model. This means they eradicate the singularity by constructing a universe that begins with a contracting phase and then bounces back to an expanding phase.  After years of continuous efforts, it is proposed that an effective field theoretic description combining the benefits of matter bounce and Ekpyrotic scenarios can give rise to a non singular cosmological model without pathologies through a Galileon-like Lagrangian. 
		\item  For RE in $\textbf{B2}$ there are two possibilities. In the first case, occurrence of bounce is realistic with usual matter. For the second case bounce occurs with exotic matter
		\item The third bouncing model is Oscillatory bounce. If the universe did experience a bounce, this may necessitate another bounce in the future. It may be possible that it is of cyclical nature. Motivated by this, we consider bounce model embedded in cyclic theories of the universe in which bounce occurs at regular intervals (oscillatory bounce). These theories do not just describe the early evolution phase of the universe, but its entire history. Consequently, recent stages such as the dark matter and dark energy domination are naturally closely tied to bounces both past and future imposing novel qualitative and quantitative constraints that can make bouncing cosmology more powerfully predictive \cite{Ijjas:2018qbo}. For example, one immediate prediction of cyclic theories is that the current dark energy dominating phase must be meta stable or slowly decaying ultimately transitioning to a state of low energy density that will initiate a contraction period. Cycling may also explain the magnitude of the dark energy density and other fundamental parameters. However, our results in Oscillatory bounce deals with the behavior of time-like geodesic congruence in the neighborhood of bouncing point using RE and FT (Focusing Theorem).
	\end{enumerate}
	As a consequence, the phenomenologies of a non singular bounce in the very early universe could be associated with the quintom scenario as inspired by the dark energy study of late time acceleration.
	Although the present study is a theoretical or rather a mathematical study of bounce, yet there are some observational aspects of the models under consideration. In this context it is worthy to refer \cite{Cai:2014bea} and \cite{Mielczarek:2010ga}.
	\begin{enumerate}
		\item For example in \cite{Cai:2014bea}, the author proposed some possible mechanisms of generating a red tilt for primordial curvature perturbations and confront its general predictions with current CMB observations. Non singular bounce that attempts to address the issue of big-bang singularity went through a series of considerable developments which led to brand new predictions of cosmological signatures, visible in many observations.
		\item From the perspective of phenomenological considerations there is ``matter bounce" scenario that gives rise to almost scale invariant power spectra of primordial perturbations and thus can fit to observations very well. For ref. see \cite{Cai:2014bea}
		\item In the paper \cite{Mielczarek:2010ga}, it was found that the Big Bounce predictions do not conflict with the observational data rather they agree with it.
	\end{enumerate}
	\newpage
	\section{Conclusion}
	The paper has examined bouncing scenario or particularly the bouncing point both geometrically and physically. General prescription of bouncing cosmologies along with examples of some peculiar bounces have been dealt with in more details.  It has been found that the bouncing point may be (i) a point of local maximum/minimum for the scale factor, (ii) a cusp in case of Super bounce, (iii) an oscillatory bouncing point, (iv) a point of inflection (w.r.t  either the curve of scale factor or the curve of Hubble parameter) or (v) a corner point. Subsequently Raychaudhuri equation has been implemented to analyse the bouncing scenario and to dictate the favourable conditions for different types of bounce to occur. Besides, the role of curvature in bouncing has been investigated and it has been found that positive curvature favors the bouncing scenario. Further it has been shown that by a suitable transformation of variable it is possible to identify the Raychaudhuri equation with the evolution equation for a classical linear Harmonic oscillator and it is found that the convergence condition can be restated in terms of the frequency of the oscillator. Further in this context, behavior of a congruence of time-like geodesics near the bouncing point (precisely in the deleted neighbourhood of the bouncing point) of an oscillatory bouncing model has been investigated using the Focusing Theorem. Moreover the Raychaudhuri equation in case of \textbf{B1} hints the existence of second order singularity at the bouncing point hence posing problem in defining higher order cosmographic parameters like $j$, $s$ etc. Therefore to define them two different equation of states have been considered namely the van der Waals EoS and Polytropic EoS as an alternative to barotropic EoS and some probable conditions have been determined to define the cosmographic parameters at least upto $q$. However Raychaudhuri equation in \textbf{B2} shows that focusing alone does not imply the formation of singularity or in other words for convergence condition there may be bouncing scenario without the existence of singularity. On the other hand, bouncing scenario is  also possible for violation of convergence condition with exotic matter. Thus extensive analysis of the Raychaudhuri equation in various bouncing models shows that the cosmological singularity (i.e, big-bang singularity) and bouncing scenario are two independent notions in cosmology and one does not imply the other generally. Behavior (continuity) of energy density and pressure for the bouncing models have also been studied using the Raychaudhuri equation. Moreover, the advantage of getting the bouncing point as a point of inflection over being cusp or corner lies in the information regarding existence and continuity of the higher order derivatives of $H$ in the neighbourhood of bouncing point including the bouncing point itself. Finally some observational aspects and challenges faced by different bouncing models to become practically relevant have been discussed. 
	\section*{Acknowledgment}
	The authors thank anonymous reviewers for suggesting insightful amendments which improved the quality of the paper. M.C thanks University Grant Commission (UGC) for providing the Junior Research Fellowship (ID:211610035684/JOINT CSIR-UGC NET JUNE-2021). S.C. thanks FIST program of DST, Department of Mathematics, JU (SR/FST/MS-II/2021/101(C)).
	
\end{document}